\documentclass[prc,superscriptaddress,unsortedaddress,twocolumn,showpacs,preprintnumbers,amsmath,amssymb,floatfix,dvipdfmx]{revtex4}
\usepackage{amsmath}
\usepackage{amssymb}
\usepackage{mathrsfs}
\usepackage{bm}
\usepackage{color}
\usepackage{graphicx}
\usepackage{braket}
\usepackage{mathrsfs}
\usepackage{ulem}

\makeatletter
\let\MYcaption\@makecaption
\makeatother
\usepackage{subcaption}
\makeatletter
\let\@makecaption\MYcaption
\makeatother

\makeatletter
\newcommand{\figcaption}[1]{\def\@captype{figure}\caption{#1}}
\newcommand{\tblcaption}[1]{\def\@captype{table}\caption{#1}}
\makeatother

\begin{document}

\title{
Investigation of contributions of the $2_2^+$ resonance in $^6$He via analysis of 
$^6$He($p$, $p'$)}

\author{Shoya Ogawa}
\email[]{s-ogawa@phys.kyushu-u.ac.jp}
\affiliation{Department of Physics, Kyushu University, Fukuoka 819-0395, Japan}

\author{Takuma Matsumoto}
\email[]{matsumoto@phys.kyushu-u.ac.jp}
\affiliation{Department of Physics, Kyushu University, Fukuoka 819-0395, Japan}

\date{\today}

\begin{abstract}
 We investigate the contribution of the $2^{+}_{2}$ resonance in $^6$He
 to observables via analysis of the $^6$He($p,p'$) reaction by
 using the continuum-discretized coupled channels method combined with the
 complex-scaling method.
 In this study, we obtain the $2^{+}_{2}$ state with the resonant energy
 2.25 MeV and the decay width 3.75 MeV
 and analyse contributions of resonances and nonresonant continuum states 
 to the cross section separately.
 It is found that the $2^{+}_{2}$ state plays an important role in
 the energy spectrum. Furthermore, contributions of nonresonant
 continuum states
 are also important to clarify the properties of the $2^{+}_{2}$ state. 
\end{abstract}

\maketitle

%
Studies on resonances have attracted much attention in many-body
quantum systems for nucleons, quarks, atoms and so on.
In nuclear physics, various resonances have been discovered and
investigated their properties, e.g. single-particle resonances, giant
resonances, and cluster resonances.
Recently, by the development of radioactive ion-beam experiments,
resonant structures of nuclei near and beyond the neutron
dripline have been intensively pursued.
Nuclei on the neutron dripline such as $^6$He, $^{11}$Li, $^{14}$Be and
$^{22}$C are known as a two-neutron halo nuclei and have a Borromean
structure in which there is no bound state in binary subsystems.
To explore resonances of such unstable nuclei, the ($p,p'$) reaction
with the inverse kinematics has been often used~\cite{Lag01,Ste02,Tan17},
and the contribution of resonances shows up as a peak structure in the 
excitation energy spectrum observed.   

For $^6$He, low-lying resonances have been discussed by
both theoretical and experimental approaches
\cite{Dan97,Pie04,Nav03,Vol05,Mic10,Hag05,Myo07,Aoy95,For14,Sin16}.
The $2^{+}_{1}$ state with a small decay width is well understood as 
the first excited state, and its contribution to cross sections 
shows up as a sharp peak.
The $2^{+}_{2}$ state, which is considered as the next lower 
state to the $2^{+}_{1}$, has also been investigated via 
structural calculations
\cite{Dan97,Pie04,Nav03,Vol05,Hag05,Myo07,Aoy95,For14,Sin16}
and experimental studies~\cite{Jan96,Mou12}.
Furthermore, the $2^{+}_{2}$ state is also considered to play
important roles in predicting the properties of some resonant states,
e.g. the $3/2^{-}_{3}$ state in $^7$He \cite{Myo07}
exists beyond the dripline
and the $3/2^{+}_{2}$, $5/2^{+}_{2}$ states in the hypernucleus
$^{7}_{\Lambda}$He \cite{Hiy15}.
Thus it is necessary to clear the property of
the $2^{+}_{2}$ state in $^6$He.
However, unfortunately,
the contribution of the $2^+_2$ state in the energy spectrum does not show
a shape structure because its resonant energy is near the right end
of the contribution of the $2^+_1$ state, and its width is rather large.
Thus, the resonant energy $(E_r)$ and the decay width $(\Gamma)$ of the
$2^+_2$ state are less clear. 
To clarify the properties of the $2_2^+$ state, it is required 
an accurate analysis of treating not only resonant contributions but also 
nonresonant ones in the energy spectrum. 

The continuum-discretized coupled-channels method (CDCC) is a reliable 
method of describing coupling effects to continuum states~\cite{Yah12}.
At first 
CDCC was a method of describing three-body breakup reactions with a 
two-body projectile, but now it is applicable to analyses of four-body 
breakup reactions, in which a projectile breaks up into three constituents 
such as two-neutron halo nuclei~\cite{Mat04,Mat06}. 
In CDCC, an energy spectrum of a breakup cross section is obtained as a 
continuous function of the excitation energy by using the smoothing procedure 
based on the complex-scaling method (CSM)~\cite{Agu71,Bal71,Aoy06,Mat10},
which is useful for searching resonances 
in many-body systems. As an advantage of the smoothing procedure,
contributions of resonances, and nonresonant continuum states in 
energy spectra are investigated separately. 
Thus the CDCC analysis combined with CSM is indispensable to investigate 
resonant contributions in energy spectra.
Recently, the energy spectrum of the $^{11}$Li($p,p'$)
reaction~\cite{Tan17} has been analyzed by the approach, and
contributions of the resonance and nonresonant continuum states in the
energy spectrum has been confirmed~\cite{Mat19}. 

In this Rapid Communication, the $2_2^+$ resonant state in $^6$He is
investigated via the CDCC analysis combined with CSM of 
$^6$He($p,p'$) reactions. In this analysis, 
the reactions are described as a $^4$He$~+~n~+~n~+~p$ four-body system,
and each resonance in  
$^6$He is obtained by CSM. The calculated elastic and inelastic 
cross sections are compared with the experimental data, 
and the effect of the $2_2^+$ state on the energy spectrum is 
discussed by excepting contributions of the $2_1^+$ state and
nonresonant continuum states.

In the $^4$He$~+~n~+~n~+~p$ four-body system,  
the Schr{\"o}dinger equation based on the multiple scattering 
is written as
\begin{eqnarray}
 \label{4body-eq}
 \left[
  K_{R}
  + \frac{A_{\rm{P}}-1}{A_{\rm{P}}}\sum_{i\in ^{6}\rm{He}}g_{i}
  + V_{C} + h - E
 \right]\Psi(\bm{\xi},\bm{R})=0,
\end{eqnarray}
where $\bm{R}$ and $\bm{\xi}$ represent the coordinates between $p$ and
the center-of-mass (c.m.) of $^6$He and the intrinsic coordinate of $^6$He,
respectively.
$K_{R}$ is a kinetic energy operator associated with $\bm{R}$, and 
$h$ is the internal Hamiltonian of $^6$He.
As the interaction between nucleons in $^6$He and $p$, $g_i$,
the Jeukenne-Lejeune-Mahaux (JLM) effective NN interaction~\cite{JLM77}
is adopted. Here it should be noted that we take into account
the antisymmetrization between nucleons in $^6$He and $p$ based on the 
Kerman--McManus--Thaler (KMT) theory~\cite{Ker59,Min17} as a factor
$(A_{\rm{P}}-1)/A_{\rm{P}}$ with the mass number $A_{\rm P}$ of $^6$He.
$V_{C}$ is the Coulomb interaction between the c.m. of $^6$He and $p$, 
that is, Coulomb breakup is neglected in the present analysis.

In CDCC, the scattering wave function with the total spin $J$ and 
its projection on $z$-axis $M$ is expanded in terms of a set of 
eigenstates $\{\Phi_{n}^I\}$ of $^6$He,  as
\begin{eqnarray}
 \Psi_{JM}(\bm{\xi},\bm{R})
  =
  \sum_{I,n,L}
  \frac{\chi_{\gamma}(K_{n},R)}{R}
  \left[
   \Phi_{n}^{I}(\bm{\xi}) \otimes i^{L}Y_{L}(\hat{\bm{R}})
  \right]_{JM} .
\end{eqnarray}
Here $I$ is the internal spin of $^6$He, and a set of $\Phi_{n}^{I}$
is generated with the Gaussian expansion method (GEM)~\cite{Hiy03},
in which $h$ is diagonalized by using Gaussian basis functions.
As a model space in the present analysis,
we take $I^{\pi}=0^{+}$, $1^{-}$ and $2^{+}$ states of $^6$He.
The orbital angular momentum regarding $R$ is $L$,  
$\gamma$ represents $\{n,I,L\}$,
and $K_n$ is the relative wave number defined by
\begin{eqnarray}
 K_n&=&\frac{\sqrt{2\mu(E-\epsilon_n^I)}}{\hbar}
\end{eqnarray}
with the reduced mass $\mu$ of the $^6$He--$p$ system and the eigen energy
$\epsilon_{n}^I$ of $\Phi_{n}^I$. 
The relative wavefunction $\chi_{\gamma}$ between the c.m. of $^6$He
and $p$ satisfies
\begin{widetext}
 \begin{eqnarray}
  \label{CDCC-eq}
   \left[
    -\frac{\hbar^{2}}{2\mu} \frac{d^{2}}{dR^{2}}
    + \frac{\hbar^{2}L(L+1)}{2\mu R^2} 
    + \frac{A_{\rm{P}}-1}{A_{\rm{P}}}U_{\gamma\gamma}(R)
    + \frac{2 e^{2}}{R}
    - (E - \varepsilon^{I}_{n})
   \right]
   \chi_{\gamma}(K_{n},R)
   =-\sum_{\gamma' \ne \gamma}
   \frac{A_{\rm{P}}-1}{A_{\rm{P}}}U_{\gamma\gamma'}(R)\chi_{\gamma'}
   (K_{n'},R).
 \end{eqnarray}
\end{widetext}
The coupling potentials $U_{\gamma\gamma'}$ between the $\gamma$ and
$\gamma'$ channels are calculated by using the folding model with the JLM
interaction, in which
a normalization factor
$N_{w}$ for the imaginary part is introduced 
by optimizing the experimental data.
Details for the calculation of $U_{\gamma\gamma'}$ are shown
in Ref.~\cite{Mat11,Mat19}.

Solving Eq.~\eqref{CDCC-eq} under the appropriate
boundary condition, we obtain a scattering $T$ matrix represented
by $T'_{nIL}$. Here it should be noted that $T'_{nIL}$ is not the actual
scattering $T$ matrix considered in the present work. 
In the KMT theory, the actual scattering $T$ matrix, $T_{nIL}$,
is defined as
\begin{eqnarray}
 T_{nIL} = \frac{A_{\rm{P}}}{A_{\rm{P}}-1} T'_{nIL}.
\end{eqnarray}
Details for the KMT theory are shown in Ref.~\cite{Ker59,Min17}.

According to the smoothing procedure based on CSM~\cite{Mat10}, 
the double differential cross section (DDX), which depends on
the internal energy $\varepsilon$ of $^6$He and the scattering angle, is
described as 
\begin{eqnarray}
 \label{DDX}
 \frac{d^{2}\sigma}{d\varepsilon d\Omega}
  =
  \frac{1}{\pi}
  {\rm Im}
  \sum_{iIL} \frac{T^{\theta}_{iIL}\tilde{T}^{\theta}_{iIL}}
  {\varepsilon-\varepsilon^{I}_{\theta,i}}
\end{eqnarray}
with
\begin{eqnarray}
 \tilde{T}^{\theta}_{iIL}
  &=&
  \sum_{n}
  \braket{\tilde{\Phi}^{I}_{\theta,i}|U(\theta)|\Phi^{I}_{n}}
  T_{nIL} ,
  \\
  T^{\theta}_{iIL}
  &=&
  \sum_{n}
  T_{nIL}
  \braket{\Phi^{I}_{n}|U^{-1}(\theta)|\tilde{\Phi}^{I}_{\theta,i}} .
\end{eqnarray}
Here $U(\theta)$ is the scaling transformation operator in CSM, and
$\Phi_{\theta,i}^{I}$ and $\varepsilon^{I}_{\theta,i}$
represent the $i$-th eigen state and energy of $^6$He, respectively.
These states calculated by using the framework combining the GEM
and CSM with the scaling angle $\theta$.
From Eq.~\eqref{DDX}, one sees that the DDX is given by an incoherent
sum of the contributions of $\Phi_{\theta,i}$. Therefore
we can distinguish between resonant and nonresonant contributions in
the DDX.

For the internal Hamiltonian of $^6$He, 
we take the Minnesota interaction~\cite{Tho77} and the KKNN
potential~\cite{Kan79} 
for the $n$-$n$ and the $n$-$^4$He interactions, respectively.
The particle exchange between valence neutrons and neutrons in $^4$He 
is treated with the orthogonality condition model~\cite{Sai69}.
Furthermore, we introduce the phenomenological
three-body potential to reproduce the energies of the ground and
$2^{+}_{1}$ states~\cite{Ajz88}. 
As the result, we obtained the ground state energy $-0.972$ MeV
and ($E_{r}$, $\Gamma$) = (0.848~MeV, 0.136~MeV) for
the $2^{+}_{1}$ state.

\begin{figure}[htbp]
 \includegraphics[scale=0.75]{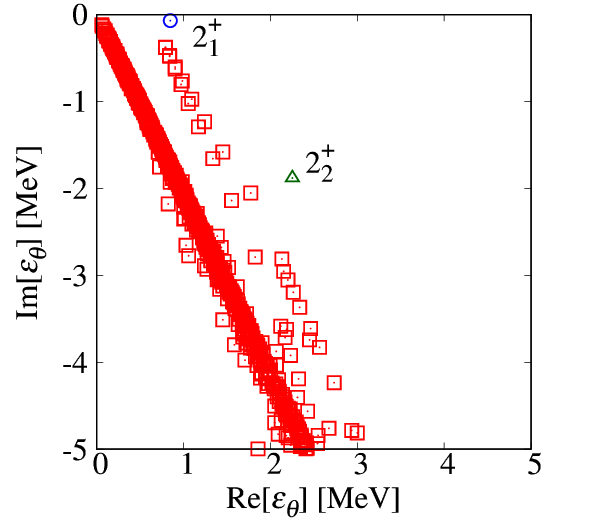}
 \caption{
 The eigen value distributions for $I^{\pi}=2^{+}$ states
 in CSM with scaling angle $\theta=32^{\circ}$.
 }
 \label{csm-ene}
\end{figure}

%
%
First, we calculate eigen energies of $^6$He by using CSM with
complex-range Gaussian functions~\cite{Oht13}, 
which are useful for obtaining resonances with large decay width.
In Fig. \ref{csm-ene}, 
the squares represent the eigen energies of nonresonant
continuum states for $I^{\pi}=2^{+}$ with the scaling angle
$\theta=32^{\circ}$.
The circle and triangle describe the eigen energies of the $2^{+}_{1}$ and
$2^{+}_{2}$ states, respectively.
It is found that the $2^{+}_{2}$ state has 
($E_{r}$,~$\Gamma$) = (2.25~MeV, 3.75~MeV), and its value
is consistent with one calculated by Myo {\it et al}~\cite{Myo07},
which provides a better description of the experimental data
as mentioned in Ref.~\cite{Mou12}.
In Table~\ref{tab:csm-ene}, all resonances obtained in the present
calculation are shown.
For the calculation of the DDX, the scaling angle is taken as 32$^\circ$ 
for the $2_2^+$ state and 15$^\circ$ for the other states. In this analysis,
the convergence of the calculated DDX has been confirmed.

\begin{figure}[tbp]
 \begin{center}
  \includegraphics[width=0.43\textwidth]
  {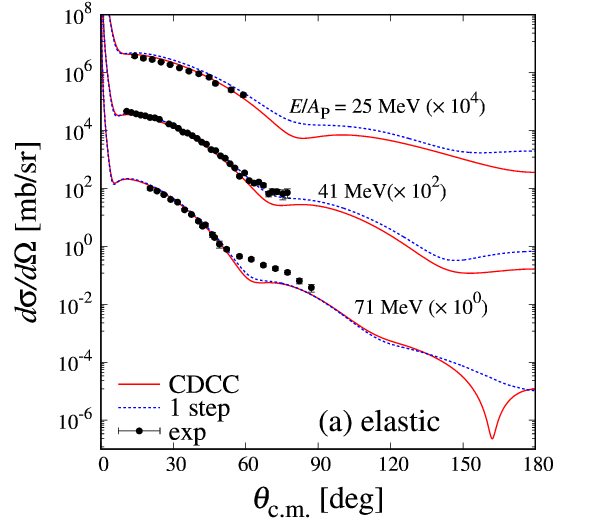}
  \includegraphics[scale=0.75]{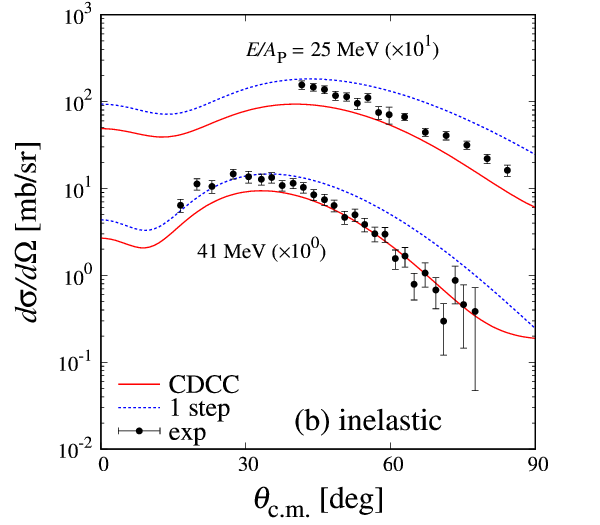}
  \caption{
  Angular distributions of the differential elastic cross section 
  at $E/A_{\rm{P}}=25$--$71$ MeV (a)
  and the differential inelastic cross section
  calculated by integrating the DDX over $\varepsilon$
  from $\varepsilon = 0$ to
  $1.5$ MeV at $E/A_{\rm{P}}=25$ and $41$ MeV (b).
  The experimental data are taken
  from Ref.~\cite{Gio05,Lag01,Kor97,Ues10,Ste02}.
  Each cross section is multiplied by the factor shown in the figure.
  }
  \label{fig2}
 \end{center}
\end{figure}

%
Next we calculate angular distributions of the elastic and inelastic
cross sections for $^6$He scattering on $p$.
Figure \ref{fig2} shows the angular distributions of the 
elastic cross section at $E/A_{\rm{P}}=25$,
$41$ and $71$ MeV~\cite{Gio05,Lag01,Kor97,Ues10} in panel
(a) and of the inelastic cross section at
$E/A_{\rm{P}}=25$
and $41$ MeV~\cite{Ste02,Lag01} in panel (b).
$\theta_{\rm c.m.}$ means the scattering angle in the center-of-mass frame.
For the inelastic cross section, the DDX is integrated over
$\varepsilon$ up to 1.5 MeV, which is the same as the experimental set
up at $E/A_{\rm{P}}=41$ MeV~\cite{Lag01}. 
The solid and dashed lines represent the results with CDCC
and those with the so-called one-step calculation, respectively.
The one-step calculation neglects multi-coupling effects in CDCC.
In the analysis, $N_w$ is taken as 0.8, and it is found that
CDCC well reproduces the experimental data for both elastic and
inelastic cross sections simultaneously.
The difference between the results of CDCC and the one-step calculation
represents coupling effects.
One sees that the effects are particularly important for
elastic cross sections at low incident energy and inelastic cross sections.
\begin{table}[htbp]
 \tblcaption{The comparing the results of the resonant energy and
 the decay width (in unit of MeV)
 with those calculated in Ref.~\cite{Myo07}.
 For our calculation, we mention $E_{r}$, the excitation energy $E_{x}$
 and $\Gamma$.}
 \label{tab:csm-ene}
 \begin{tabular}{cp{3em} cp{8em} cp{8em} cp{8em} cp{8em} cp{8em}}
  \hline
  &~ &\multicolumn{3}{c}{present results}
  &\multicolumn{2}{c}{~~~~Ref.~\cite{Myo07}}
  \\
  \hline
  &\multicolumn{1}{c}{$I^{\pi}$}
  &\multicolumn{1}{c}{~~$E_{r}$}
  &\multicolumn{1}{c}{~~~~$E_{x}$}          
  &\multicolumn{1}{c}{~~~~$\Gamma$}
  &\multicolumn{1}{c}{~~~~~~$E_{x}$}
  &\multicolumn{1}{c}{~~~~$\Gamma$} \\
  \hline
  &\multicolumn{1}{c}{$2^{+}_{1}$}
  &\multicolumn{1}{c}{~~0.848}
  &\multicolumn{1}{c}{~~~~1.83}
  &\multicolumn{1}{c}{~~~~0.136}
  &\multicolumn{1}{c}{~~~~~~1.8}
  &\multicolumn{1}{c}{~~~~0.1}
  \\
  &\multicolumn{1}{c}{$2^{+}_{2}$}
  &\multicolumn{1}{c}{~~2.25} 
  &\multicolumn{1}{c}{~~~~3.23}
  &\multicolumn{1}{c}{~~~~3.75}
  &\multicolumn{1}{c}{~~~~~~3.5}
  &\multicolumn{1}{c}{~~~~3.9}
  \\
  &\multicolumn{1}{c}{$0^{+}_{2}$}
  &\multicolumn{1}{c}{~~3.70} 
  &\multicolumn{1}{c}{~~~~4.68}
  &\multicolumn{1}{c}{~~~~7.13}
  &\multicolumn{1}{c}{~~~~~~4.9}
  &\multicolumn{1}{c}{~~~~8.8}
  \\
  &\multicolumn{1}{c}{$1^{-}_{1}$}
  &\multicolumn{1}{c}{~~4.42}
  &\multicolumn{1}{c}{~~~~5.39}
  &\multicolumn{1}{c}{~~~~4.82}
  \\
  &\multicolumn{1}{c}{$1^{-}_{2}$}
  &\multicolumn{1}{c}{~~4.51}
  &\multicolumn{1}{c}{~~~~5.48}
  &\multicolumn{1}{c}{~~~~8.23}
  \\
  \hline
 \end{tabular}       
\end{table}

\begin{figure}[htbp]
 \begin{center}
  \includegraphics[scale=0.75]{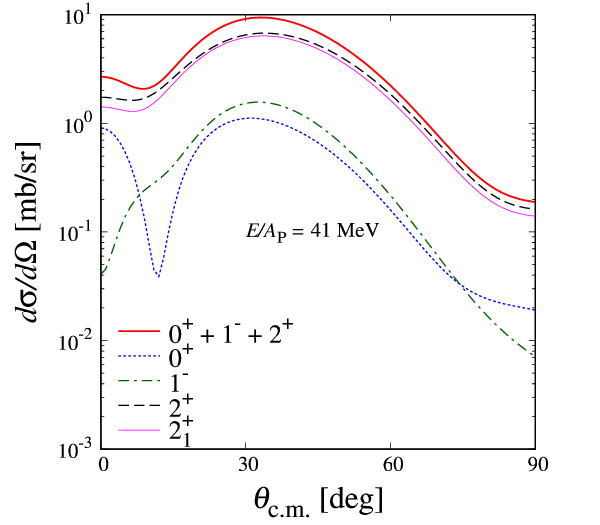}
  \caption{
  Angular distributions of the differential cross section
  calculated by the same way as Fig \ref{fig2}(b)
  in the case of $^6$He + $p$ inelastic scattering 
  at $E/A_{\rm{P}}=41$ MeV.
  }
  \label{fig4}
 \end{center}
\end{figure}

For the inelastic cross section, we investigate the contribution of
the $2_1^+$ state.
Figure \ref{fig4} shows the inelastic cross section at 
$E/A_{\rm{P}}=41$ MeV for each spin-parity state of $^6$He.
The thick solid line is the same as the result of CDCC in Fig.~\ref{fig2}(b).
The dotted, dot-dashed, dashed, and thin solid lines denote 
the contributions
for $I^{\pi}=0^{+}$, $1^{-}$, $2^{+}$, and of
the $2^{+}_{1}$ state, respectively.
It is found that the contribution for $I^\pi=2^{+}$ is dominant
and mainly comes from the $2^+_1$ state.
Further, contributions for $I^{\pi}=0^{+}$ and $1^{-}$ are
not negligible and come from non resonant continuum states
because resonances for $I^{\pi}=0^{+}$ and $1^{-}$ do not exist in 
$\varepsilon \leq 1.5$ MeV.
This result shows that the experimental data shown 
in Fig.~\ref{fig2}(b) includes 
not only the contribution of the $2_1^+$ state but also 
contributions of nonresonant continuum states. 
In the present calculation, the nonresonant contributions 
account for about 30\% of the total. 
%

\begin{figure}[tp]
 \begin{center}
  \includegraphics[scale=0.75]{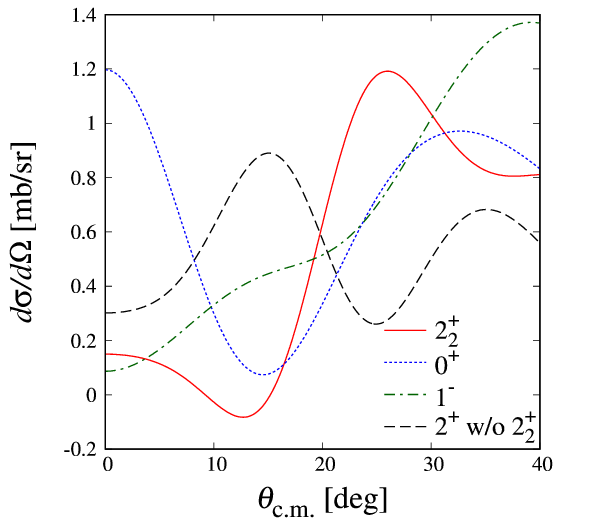}
  \caption{
  Angular distributions of the differential cross section 
  calculated by integrating the DDX up from $\varepsilon = 1.5$ to 3 MeV.
  }
  \label{fig5}
 \end{center}
\end{figure}
\begin{figure}[tbp]
 \begin{center}
  \includegraphics[scale=0.75]{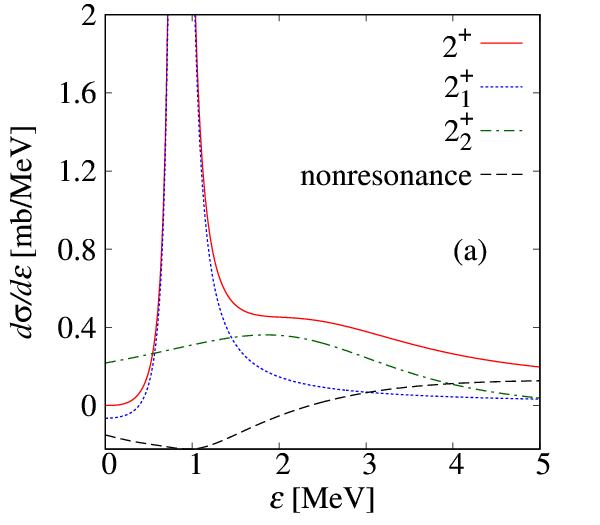}
  \includegraphics[scale=0.75]{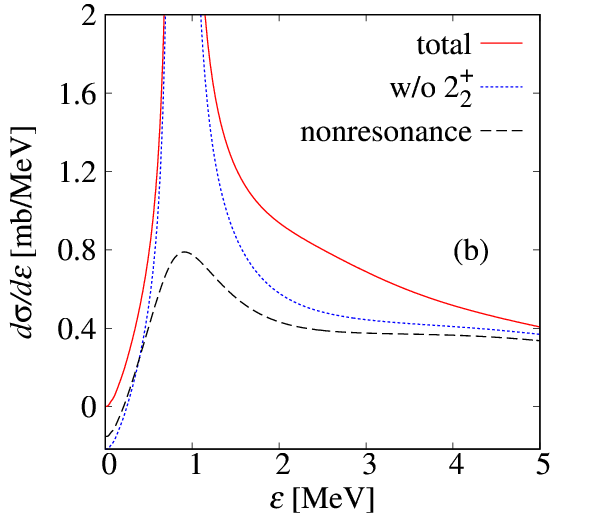}  
  \caption{
  Energy spectra of the breakup cross section calculated by
  integrating the DDX up from $\theta_{\rm c.m.} = 20^{\circ}$ to
  $30^{\circ}$ for $I^\pi=2^+$ (a) and the sum of all spin-parity states
  (b).
  }
  \label{fig6}
 \end{center}
\end{figure}
Finally, we discuss the contribution of the $2^{+}_{2}$ state
to energy spectra of the breakup cross section
in the case of $E/A_{\rm{P}}=41$ MeV.
As mentioned above, for the angular distribution of 
the DDX integrated over $\varepsilon$ up to 1.5 MeV, the contribution of 
the $2_2^+$ state is negligible. Therefore we investigate the 
angular distribution of the DDX integrated over $\epsilon$ from 1.5 MeV to 
3 MeV as shown in Fig.~\ref{fig5}.
The solid line represents the contribution of the $2^{+}_{2}$ state.
The dotted, dot-dashed, dashed lines describe the contributions for
$I^{\pi}=0^{+}$, $1^{-}$ and
$I^\pi=2^{+}$ without the $2^{+}_{2}$ state.
One sees that the contribution of the $2^{+}_{2}$ state
is more important than the other contributions in  
$20^{\circ} \leq \theta_{\rm c.m.} \leq 30^{\circ}$.
Then we focus on the energy spectrum around the scattering angle region.

Figure \ref{fig6}(a) shows the energy spectrum 
for $I^{\pi}=2^{+}$ calculated by integrating 
the DDX over $\theta_{\rm c.m.}$ from $20^{\circ}$ to $30^{\circ}$. 
The solid line represents the total contribution for $I^{\pi}=2^{+}$.
The dotted, dot-dashed, and dashed lines denote the contributions of
the $2^{+}_{1}$, $2^{+}_{2}$ and nonresonant continuum states, respectively.
The strong peak around $\varepsilon=1$ MeV comes from the $2^{+}_{1}$ state.
Meanwhile the contribution of the $2^{+}_{2}$ state
shows up as a shoulder peak and is dominant around $\varepsilon=2$ MeV.
Thus in terms of the cross section for only $I^\pi=2^+$, the contribution
of the $2^+_2$ state is rather visible in the energy spectrum.

However it is difficult to extract a component of a specific spin-parity 
state from the experimental data. Therefore we have to confirm how to see the
contribution of the $2^+_2$ state in the cross section including components 
from all spin-parity states of $^6$He.
In Fig.~\ref{fig6}(b), the solid line represents the energy spectrum 
with the all components in the present model space, and the dotted 
line corresponds to the result without the 
$2_2^+$ state.
If there was not the $2^+_2$ state, the shape of the cross section would
be shown by the dotted line.
Although a peak structure from the $2^+_2$ state does not exist,
an increase of the cross section around $\varepsilon=2$ MeV is found by
comparing the solid line with the dotted line.
The similar increase of the cross section can be confirmed in the measured
energy spectrum from $^8$He($p,t$)$^6$He reaction in Ref.~\cite{Mou12}. 
Therefore we conclude that the increase of the cross section around
$\varepsilon=2$ MeV indicates the existence of the $2_2^+$ state.
Here it should be noted that the low-lying energy spectrum includes 
contributions of nonresonant continuum states shown by the dotted line.
It indicates that one fails to obtain properly the resonant energy
and decay width of the $2^+_2$ state if the contributions of
nonresonant continuum states are neglected.
To clarify the above indication,
we fit the breakup cross section up to $\varepsilon$ = 3 and 4 MeV
by using the Breit--Wigner functions,
\begin{eqnarray}
 f(\varepsilon)=
  \sum_{i=1}^{2}
  A^{i}
  \frac{\Gamma^{i}/2}{(\varepsilon-E^{i}_{r})^2+(\Gamma^{i}/2)^2} ,
\end{eqnarray}
where $A^{i},~E^{i}_{r}$ and $\Gamma^{i}$ are free parameters, and
$i$ = 1 and 2 represent the $2^+_1$ and $2^+_2$ states, respectively.
This Breit--Wigner parametrization does not take into account effects of
nonresonant continuum states.
In Table \ref{tab:fit}, the obtained resonant energies and decay widths
by fitting are shown.
For the $2^{+}_{1}$ state,
$E_{r}$ and $\Gamma$ are consistent with the result of the CSM, and
nonresonant effects are negligible.
Meanwhile, $E_{r}$ and $\Gamma$ of the $2^{+}_{2}$ state
are smaller than those calculated in the CSM because nonresonant effects
are neglected.
Furthermore the properties of the $2^{+}_{2}$ state thus obtained
get close to those determined in $^8$He($p,t$)$^6$He
reaction \cite{Mou12}.
This implies that the higher resonant energy and broader decay width
can be obtained if a detail analysis taken into account effects
of nonresonant continuum states for $^8$He($p,t$)$^6$He reaction is
performed.
Therefore, in order to determine the resonant energy and decay width of
the $2^+_2$ state, an accurate analysis of treating not only resonant
contributions but also nonresonant ones is highly required.

\begin{table}[htbp]
 \tblcaption{The comparing the results of the calculated resonant energy and
 the decay width (in unit of MeV) in the CSM  with those obtained by fitting
 the breakup cross section up to $\varepsilon$ = 3 and 4 MeV.
 }
 \label{tab:fit}
 \begin{tabular}{cp{1em} cp{8em} cp{8em} cp{8em} cp{8em} cp{8em} cp{8em}}
  \hline
  &~
  &\multicolumn{2}{c}{CSM}
  &\multicolumn{2}{c}{~~ Fit [0--3 MeV]}
  &\multicolumn{2}{c}{~~ Fit [0--4 MeV]}
  \\
  \hline
  &\multicolumn{1}{c}{$I^{\pi}$}
  &\multicolumn{1}{c}{~~$E_{r}$}
  &\multicolumn{1}{c}{~~~~$\Gamma$}
  &\multicolumn{1}{c}{~~~~~~$E_{r}$}
  &\multicolumn{1}{c}{~~~~$\Gamma$}
  &\multicolumn{1}{c}{~~~~~~$E_{r}$}
  &\multicolumn{1}{c}{~~~~$\Gamma$} \\
  \hline
  &\multicolumn{1}{c}{$2^{+}_{1}$}
  &\multicolumn{1}{c}{~~0.848}
  &\multicolumn{1}{c}{~~~~0.136}
  &\multicolumn{1}{c}{~~~~~~0.855}
  &\multicolumn{1}{c}{~~~~0.144}
  &\multicolumn{1}{c}{~~~~~~0.855}
  &\multicolumn{1}{c}{~~~~0.144}
  \\
  &\multicolumn{1}{c}{$2^{+}_{2}$}
  &\multicolumn{1}{c}{~~2.25} 
  &\multicolumn{1}{c}{~~~~3.75}
  &\multicolumn{1}{c}{~~~~~~1.71} 
  &\multicolumn{1}{c}{~~~~1.97}
  &\multicolumn{1}{c}{~~~~~~1.92}
  &\multicolumn{1}{c}{~~~~2.72}
  \\
  \hline
 \end{tabular}       
\end{table}
%

%
%

To summarize, we investigated the contribution of the $2^{+}_{2}$ state
in $^6$He to the breakup cross section via the CDCC analysis combined with
CSM of $^6$He($p,p'$) reactions.
As the result of CSM, we obtained the resonant energy and decay width of
the $2^{+}_{2}$ state, which are consistent with those in the previous study.
In the analysis of $^6$He($p,p'$) reactions,
we calculated the angular distributions of the elastic and inelastic
scattering and confirmed importance of coupling effects.
For the inelastic cross section, which the DDX is integrated over
$\varepsilon$ up to 1.5 MeV, it is found that not only the
$2^{+}_{1}$ state but also nonresonant continuum states contribute
substantially to the cross section.

Furthermore, we calculated the breakup cross section by integrating
the DDX from $\theta_{\rm c.m.}=20^{\circ}$ to $30^{\circ}$ to 
investigate the contribution of the $2^{+}_{2}$ state to the energy
spectrum. 
As the result, the shoulder peak due to the $2^{+}_{2}$ state appears in
the component for $I^{\pi}=2^{+}$ around $\varepsilon=2$ MeV,
and the effect from the existence of the $2^{+}_{2}$ state is enhanced
in the total components in $\varepsilon=2\mbox{--}3$. 
Moreover it is found that the contribution of nonresonant continuum states
to the breakup cross section is also important.
Indeed, the nonresonant contribution affects on the resonant energy and
decay width estimated from a fitting of cross sections.
Thus an accurate analysis
of treating both resonances and nonresonant continuum states is highly
required to clarify the properties of the $2^{+}_{2}$ state.
To discuss in more details, new experimental data are also desired.

\section*{Acknowledgements} 
This work is supported in part by Grant-in-Aid for Scientific Research
(No.\ JP18K03650)
from Japan Society for the Promotion of Science (JSPS). 


\end{document}